\documentclass{article}
\usepackage{spconf,amsmath,graphicx}
\usepackage{psfrag}

\usepackage[absolute]{textpos}
\newcommand{\copyrightstatement}{
    \begin{textblock}{15}(0.5,0.3)    
         \noindent
         \centering
         \textblockcolour{white}
         \footnotesize
         \copyright 2017 IEEE. Personal use of this material is permitted. Permission from IEEE must be obtained for all other uses, in any current or future media, including reprinting/republishing this material for advertising or promotional purposes, creating new collective works, for resale or redistribution to servers or lists, or reuse of any copyrighted component of this work in other works
    \end{textblock}
}

\title{Video Decoding Energy Estimation Using Processor Events}
\name{Christian Herglotz and Andr\'e Kaup}
\address{Multimedia Communications and Signal Processing\\
Friedrich-Alexander University Erlangen-N\"urnberg,
Cauerstr. 7, 91058 Erlangen, Germany\\ Email: \{ \ christian.herglotz, andre.kaup\ \}\ @~FAU.de}
\begin{document}
\copyrightstatement
%
\maketitle
\begin{abstract}
In this paper, we show that processor events like instruction counts or cache misses can be used to accurately estimate the processing energy of software video decoders. Therefore, we perform energy measurements on an ARM-based evaluation platform and count processor level events using a dedicated profiling software. Measurements are performed for various codecs and decoder implementations to prove the general viability of our observations. Using the estimation method proposed in this paper, the true decoding energy for various recent video coding standards including HEVC and VP9 can be estimated with a mean estimation error that is smaller than $6\%$. 
\end{abstract}
\section{Introduction}
\label{sec:intro}
During the past decades, research in the field of video coding achieved great increases in compression efficiency such that nowadays, high quality video communication is feasible whenever a wireless network like wireless LAN (WiFi) or GSM is available. Concurrently, the complexity of the corresponding decoder and encoder implementations has increased dramatically which is caused by more sophisticated compression tools. As nowadays, popular video communication services like YouTube or Netflix are often used on portable devices like smartphones or tablet PCs, the operating time of these devices, which is limited by the battery capacity, draws more and more attention. 

As a consequence, to achieve an extended operating time without having to augment the battery capacity, a high amount of research aims at reducing the processing energy of algorithms. 
E.g., an optimization framework for intra coded HEVC bit streams requiring less processing energy on the decoder side was proposed in \cite{Herglotz16a}. As real energy measurements are complex and costly, many works rely on processor events to approximate the decoding energy. E.g., Mallikarachchi et al. propose using the central processing unit (CPU) instruction counts $c_\mathrm{CPU}$ as a linear predictor for the energy. Assuming that the switching capacitance of the processor $C$ is constant for each instruction, the energy can be calculated by 
\begin{equation}
\label{eq:Ecap}
\hat E = C\cdot V^2 \cdot c_\mathrm{CPU}, 
\end{equation}
where $V$ is the supply voltage. An equivalent formulation was also used by He et al. for the encoder \cite{He05}. Likewise, Ren et al. used the processor events instruction counts, level 1 data cache misses, and hardware interrupts  to estimate the processing energy and achieved small estimation errors when employing a multivariate adaptive regression spline (MARS) model \cite{Ren14}. 

In a similar direction, several works directly aim at reducing the processing complexity to be able to keep processor constraints. In this direction, van der Schaar et al. proposed using a complexity model based on three types of processor instructions: add, multiply, and assign \cite{vdSchaar05}. Considering the processing time, Li et al. targeted the number of frames per second by using a complexity controlled encoder \cite{Li11}. 

Unfortunately, all these works do not provide experiments showing the general viability of \eqref{eq:Ecap}. E.g., the results shown by Ren et al. only evaluate the estimation error for a single decoder solution. Likewise, such a relation was examined in \cite{Herglotz15b} only for an HEVC decoder (HM \cite{HM}) on a SPARC processor. 
In this work, we examine the relation between the decoder's processing energy and various processor events that can be obtained using a standard, open source profiling tool (Valgrind \cite{valgrind}). In comparison to earlier work, this work provides the following contributions: first, the model using processor events and second, the evaluation using different codecs.  


The rest of the paper is organized as follows. Section \ref{sec:meas} introduces our measurement setup and the profiling tool used for counting the processor events. Then, Section \ref{sec:corr} discusses the correlation between the measured energy and the processor events. Finally, Section \ref{sec:model} proposes a simple calculus that can be used to accurately estimate the decoding energy depending on a small subset of 
selected processor events.

\section{Measurement Setup}
\label{sec:meas}

In this section, we present the evaluation board and the measurement setup that was used to determine the decoding energy. Furthermore, we briefly introduce the profiling tool. 

\subsection{Hardware}
\label{secsec:panda}

To prove the relation between processor instructions and decoding energy we use a pandaboard \cite{Panda}. It is an evaluation platform equipped with an OMAP4430 processor that includes an ARM Cortex-A9 dual core processor which is a typical hardware for portable devices \cite{Herglotz18}. As an operating system, we use Ubuntu 12.04 and for energy measurements, we restrict the runlevel to level $1$ to avoid interference from background processes. For remote control, we make use of the serial port. 

We show the general viability of our observations by testing the decoding process of four different codecs (namely H.263, H.264/AVC, HEVC, and VP9) with three different decoder implementations. The implementations are FFmpeg \cite{FFmpeg} which can be used for all codecs, libde265 \cite{libde} for HEVC, and TMN-2.0 \cite{TMN-2.0} for H.263. To have a large and meaningful set of test sequences, we use the same bit stream set that was already used in \cite{Herglotz16c} in which all 8-bit sequences of each class of the HEVC common test conditions \cite{Bossen13} are coded each with at least $4$ QPs. Additionally, various encoder configurations were used such that the influence of the memory (which, due to a higher amount of memory accesses, is much higher in inter prediction than in intra prediction) can be analyzed. Furthermore, to account for the fact that H.263 only allows CIF and QCIF format, the set of test sequences is extended by $10$ corresponding sequences coded with $6$ QPs. A summary of the properties of these bit streams that is taken from \cite{Herglotz16c} is shown in Table \ref{tab:config}
\begin{table}[t]
\renewcommand{\arraystretch}{1.3}
\caption{Software and configurations for encoding the evaluation bit streams. The last row in each column denotes the total number of tested bit streams. To obtain a sufficiently large test set for H.263, different parts of the input sequences were coded. The table is taken from \cite{Herglotz16c}. }
\label{tab:config}
\vspace{-.5cm}
\begin{center}
\begin{tabular}{l|c|c}
\hline
 & H.263 & VP9  \\
 \hline
 Encoder & TMN-2.0 \cite{TMN-2.0} & libvpx \cite{libvpx} \\
 Configurations & Frames $1$ to $N$ & One-pass coding \\
  & Frames $N+1$ to $2N$ & Two-pass coding \\
  QPs & $1$, $3$, $7$, $12$, $23$, $30$ & $5$, $20$, $44$, $59$\\
  \# Bit streams & $120$ & $272$ \\
  \hline \hline
  & H.264 & HEVC \\
  \hline
  Encoder & JM-18.4 \cite{JM} & HM-16.4 \cite{HM} \\
  Configurations & baseline & intra \\
  & main & lowdelay \\
  & extended & lowdelay\_P \\
  & & randomaccess \\
  QPs & $12$, $22$, $32$, $42$ & $10$, $20$, $30$, $40$\\
  \# Bit streams & $408$ & $544$ \\
  \hline
\end{tabular}
\end{center}
\end{table}

To measure the energy consumption, we use the test setup used in \cite{Herglotz18}. In this setup, we measure the energy consumption of the complete board using the voltage across and the current through the main supply jack. As a power meter, we use the high precision ZES Zimmer LMG95 which is able to internally integrate the power such that energy values are directly returned by this device. Finally, we perform an additional measurement to determine the energy consumed in idle mode (which is the offset energy) which is then subtracted from the complete energy. Hence, the pure processing energy of the decoding process $E_\mathrm{dec}$ can be obtained.

\subsection{Processor Profiling}
\label{secsec:profile}
To analyze processor events, we use the open source Valgrind framework \cite{valgrind} that supports a high amount of different processor types. It takes a desired process (in our case the decoding process) and analyzes the complete process for instructions, memory accesses, memory leaks, cache misses, and many more. Unfortunately, this instrumentation slows the original process down by a factor of $5-100$ \cite{valgrind} such that on the one hand, the application of this tool is very simple and straightforward, but on the other hand, it is highly time consuming. Note that for energy measurements, the pure decoding process without profiling was measured. 

For our work, we use the tool ''cachegrind`` that counts the following processor events: 
\begin{itemize}
\item Instruction fetches $I$: The actual amount of instructions the processor executes,
\item Data reads $R$: The number of memory accesses performed by the processor to read data,
\item Data writes $W$: The number of memory writes performed by the processor.
\end{itemize}
Furthermore, each of these numbers is further specified by the use of the cache. Hence, the following values are also given: 
\begin{itemize}
\item Reference (which is the total number of instructions and data reads and writes), 
\item L1 cache misses (which is the number of cache misses that occur on the level 1 cache)
\item LL cache misses (which is the number of cache misses that occur on the last level cache). 
\end{itemize}
The last level cache corresponds to the highest level of the cache that is available on the current processor. For the ARM Cortex-A9, this is the level 2 cache. To give an example on the relation between these values, the number of L1-cache hits can be calculated by subtracting the L1 misses from the reference. In equations, these specifications will be denoted by the subscripts r, L1, and LL such that, e.g., $R_{L1}$ denotes the number of data reads where a level-1 cache miss occurred. 
All in all, we consider $3\times 3 = 9$ types of processor events. 

\subsection{Decoding Time}
\label{secsec:time}
To have another reference, we take the decoding time 
as a reference for energy estimation. In \cite{Herglotz15a} it was shown that it can also be used to estimate the decoding energy. In this work, for FFmpeg and TMN-2.0 we use the user time provided in the linux \texttt{time}-function, and for libde265 the C++ \texttt{clock}-function. Note that due to the precision of the output format (the seconds are quantized to $0.01$ms), the instruction counts shown above are more accurate for very short processes.

\section{Correlation}
\label{sec:corr}
\begin{figure}
\centering
\psfrag{000}[r][l]{$0$}
\psfrag{001}[r][l]{$1$}
\psfrag{002}[r][l]{$2$}
\psfrag{003}[r][l]{$3$}
\psfrag{004}[r][l]{$4$}
\psfrag{005}[r][l]{$5$}
\psfrag{006}[r][r]{$0$}
\psfrag{007}[r][r]{$10$}
\psfrag{008}[r][r]{$20$}
\psfrag{009}[r][r]{$30$}
\psfrag{010}[l][c]{$I_\mathrm{LL} / 10^6$}
\psfrag{011}[l][c]{$I_\mathrm{r} / 10^{10}$}
\psfrag{012}[b][c]{Decoding energy [J]}
\psfrag{013}[t][c]{Processor count}
\includegraphics[width=0.5\textwidth]{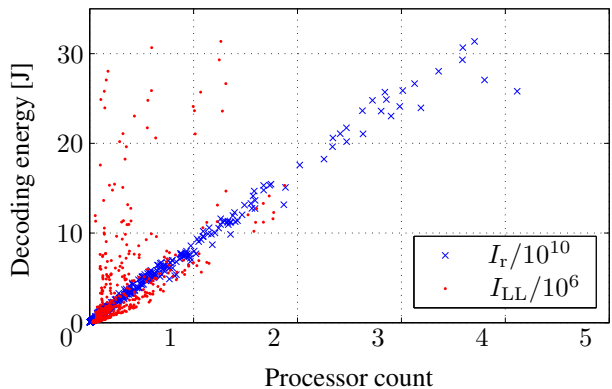}
\caption{Measured decoding energy over two types of instruction counts (HEVC decoding with FFmpeg). Each marker corresponds to the decoding process of a single bit stream. }
\label{fig:instr}
\end{figure}

In this section, we show the relation between the processor events and the decoding energy in diagrams and calculate correlation coefficients. Therefore, at first, we consider a single decoder solution (HEVC decoding with the FFmpeg software) and plot two selected processor counts and the decoding energy in Fig. \ref{fig:instr}. 

We can see that the decoding energy is highly correlated to the number of processor instructions $I_\mathrm{r}$. In contrast, the number of instructions causing a last-level cache miss $I_\mathrm{LL}$ only has a low correlation. However, we can say that all points are located above a certain threshold line such that for a fixed $I_\mathrm{LL}$, a minimum decoding energy is required. 

To express the correlation in numbers, we calculate Pearson's correlation coefficient $c$ as proposed by Bendat et al. in \cite{Bendat71}. In this calculus, $\left|c\right|=1$ means that we have a perfect linear correlation and $c=0$ means that there is no linear correlation. The correlation coefficients between the decoding energy and all considered types of instructions are listed in Table \ref{tab:corrCoeff}. 
\begin{table}[t]
\renewcommand{\arraystretch}{1.3}
\caption{Correlation coefficients between the measured decoding energy and the considered processor events (HEVC decoding with FFmpeg). }
\label{tab:corrCoeff}
\vspace{-.2cm}
\begin{center}
\begin{tabular}{l|r r r}
\hline
 & $I$ & $R$ & $W$ \\
 \hline
r  & $0.994$ & $0.995$ & $0.992$\\
L1 & $0.960$ & $0.935$ & $0.792$\\
LL & $0.553$ & $0.737$ & $0.776$\\
  \hline
\end{tabular}
\end{center}
\end{table}

We can see that the highest correlations are obtained for the number of instructions ($I_\mathrm{r}$) and the data reads ($R_\mathrm{r}$). The last-level instruction misses that are plotted in Fig. \ref{fig:instr} show the lowest correlation. For the other codecs, a similar observation can be made where in most cases, the correlation for the instructions $I_\mathrm{r}$ is slightly higher than the correlation for the data reads $R_\mathrm{r}$. 

\begin{figure}
\centering
\psfrag{000}[r][l]{}
\psfrag{001}[r][l]{$0$}
\psfrag{002}[r][l]{$0.5$}
\psfrag{003}[r][l]{$1$}
\psfrag{004}[r][l]{$1.5$}
\psfrag{005}[r][l]{$2$}
\psfrag{006}[r][r]{$0$}
\psfrag{007}[r][r]{$10$}
\psfrag{008}[r][r]{$20$}
\psfrag{009}[r][r]{$30$}
\psfrag{010}[l][c]{VP9}
\psfrag{011}[l][c]{H.264/AVC}
\psfrag{012}[l][c]{HEVC}
\psfrag{013}[b][c]{Decoding energy [J]}
\psfrag{014}[t][c]{Data reads $R_\mathrm{r}/10^{10}$}
\includegraphics[width=0.5\textwidth]{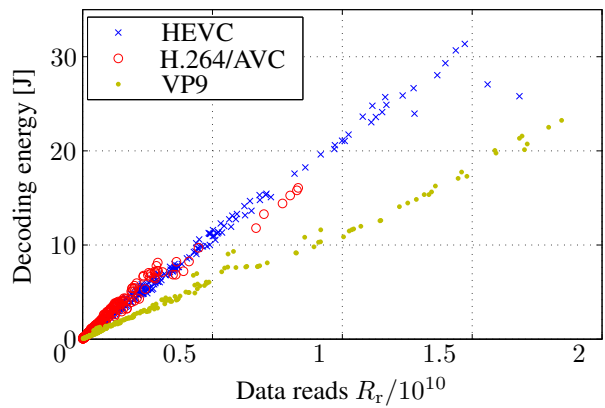}
\caption{Measured decoding energy over the number of data reads for various codecs (FFmpeg decoding). Each marker corresponds to the decoding process of a single bit stream. }
\label{fig:codecs}
\end{figure}

Finally, we would like to show that such correlations can depend on the used codec. In Fig. \ref{fig:codecs} we show the relation between data reads $R_\mathrm{r}$ and the decoding energy for three different codecs. 
We can see that a high correlation occurs for each codec (larger than $0.986$), but taking all codecs as one set into account reduces the linear correlation significantly to $0.944$.

\section{Energy Modeling}
\label{sec:model}
Finally, we show how processor events can be used to model the energy consumption. We take a purely linear approach with 
\begin{equation}
\hat E = \sum_{\forall \mathrm{PE}} n_\mathrm{PE}\cdot e_\mathrm{PE}, 
\end{equation}
where $\mathrm{PE}\in \{I_\mathrm{r}, I_\mathrm{L1}, I_\mathrm{LL}, R_\mathrm{r},...,W_\mathrm{LL}\}$ is the considered set of processor events, and $n_\mathrm{PE}$ is the number of occurrences of the respective processor event. $e_\mathrm{PE}$ can be interpreted as the mean energy required to execute a certain PE which, in this work, is assumed to be constant. Such a simple modeling function was, e.g., also used in \cite{Herglotz15b}.   

As a benchmark, we additionally model the energy using the decoding time based model that is proposed in \cite{Herglotz15a} and a bit stream feature based model that is proposed in \cite{Herglotz16c}. We calculate the estimation error by 
\begin{equation}
\bar \varepsilon = \frac{1}{M}\sum_{m=1}^M \frac{\left| \hat E_m - E_m\right|}{E_m}, 
\end{equation}
where $M$ is the complete number of bit streams, $m$ the bit stream index, $\hat E_m$ the estimated energy, and $E_m$ the measured energy of the $m$-th bit stream. 

To train the parameters $e_\mathrm{PE}$, we use a least squares fitting algorithm and perform a 10-fold cross validation to obtain realistic estimation errors (cf. \cite{Herglotz18}). We consider three cases: estimation using a single PE (where we choose instruction fetches $I_\mathrm{r}$ as they show the highest correlation to the decoding energy), estimation using all $9$ PEs, and estimation using a subset of $4$ PEs that reaches acceptable estimation errors. The four chosen PEs ($I_\mathrm{r}$, $I_\mathrm{LL}$, $W_\mathrm{r}$, and $W_\mathrm{LL}$) are carefully selected and show that the energy consumption is mainly caused by the complete number of instructions and data writes and the corresponding last-level cache misses. 

Table \ref{tab:estError} summarizes the estimation errors for all considered codecs and models. 
\begin{table}[t]
\renewcommand{\arraystretch}{1.3}
\caption{Energy estimation errors $\bar \varepsilon$ for various decoders and estimation models. For HEVC, H.264, H.263, and VP9, the FFmpeg decoder was used. libde and TMN are decoders for HEVC and H.263, respectively. In the last row, we give the estimation error when evaluating all codecs in a single set. In this case, the feature based model (that is codec dependent) cannot be evaluated.  }
\label{tab:estError}
\vspace{-.2cm}
\begin{center}
\begin{tabular}{l|r r r r r}
\hline
$\bar\varepsilon$ & $1$ PE & $4$ PE  &  $9$ PE & \cite{Herglotz15a} & \cite{Herglotz16c} \\
 \hline
HEVC  & $11.78\%$ & $3.67\%$ & $3.23\%$& $8.19\%$ & $5.27\%$ \\
\, libde & $6.27\%$ & $4.26\%$ & $4.0\%$ & $1.31\%$ & $3.18\%$\\
H.264 & $23.04\%$ & $4.13\%$ & $3.38\%$ & $4.25\%$ & $6.41\%$ \\
H.263 & $26.32\% $ & $5.86\%$ & $3.91\%$ & $4.35\%$ & $2.51\%$  \\
\, TMN & $18.64\%$ & $1.59\%$ & $0.64\%$ & $9.61\%$ & $0.77\%$ \\
VP9 & $9.14\%$ & $2.64\%$ & $2.48\%$ & $1.98\%$ & $5.11\%$ \\
  \hline
  All & $24.36\%$ & $11.82\%$ & $8.78\%$ & $7.02\%$& - \\
  \hline
\end{tabular}
\end{center}
\end{table}
E.g., the first value ($\bar \varepsilon = 11.78\%$) means that if we only use one PE to model the energy consumption of the HEVC decoder in the FFmpeg framework, then we expect an average estimation error of $11.78\%$. 

Disregarding the last row, we can see that in many cases, considering a single PE is not sufficient as estimation errors can be higher than $20\%$. Using all nine considered PEs, the estimation error drops below $4\%$. 
Using the subset of $4$ PEs, we obtain errors below $6\%$ such that choosing these four PEs is a good trade-off. 

The second last column shows the estimation errors when using the decoding-time based model \cite{Herglotz15a}. 
We can see that in comparison to using a single PE, the decoding time is always better suited for decoding energy estimation. In comparison to $4$ PEs and $9$ PEs, the decoding time shows similar estimation errors. 
Strikingly, the error for the TMN software is relatively large. This can be explained by 
 the quantized output of the timing function which causes a high error to the very short decoding processes of the TMN software. 
 However, for sufficiently long decoding processes, the decoding time is slightly better suited for decoding energy estimation than the processor events. Considering the last column that shows estimation errors using a bit stream feature based model \cite{Herglotz16c}, we can see that the errors are in the same range as using $4$ processor events. 

Finally, we consider all codecs and decoders summarized in a single set in the last row (All). I.e. that we do not distinguish between the codecs and the decoder software in model training and validation. For the feature based model \cite{Herglotz16c}, the estimation error cannot be evaluated because the bit stream features are codec specific (which is not the case for processor events and decoding time as, e.g., the number of instructions can be determined for any decoding process).  We can see that the overall estimation error increases significantly for all considered models. This can be explained by the different characteristics of the implementations that differ in terms of memory management and processor instruction types used. 

The modeling approach returns mean specific energies that can be interpreted as the energy cost that occurs if the corresponding processor event occurs once. For the model using $4$ PEs that is trained for all decoders, these values are $e_{I_\mathrm{r}}=0.47\mathrm{nJ}$, $e_{I_\mathrm{LL}}=0.43\mu\mathrm{J}$, $e_{W_\mathrm{r}}=1.5\mathrm{nJ}$, and $e_{W_\mathrm{LL}}=0.16\mu\mathrm{J}$.

\section{Conclusions}
In this paper, we showed that processor events can be used to accurately estimate the processing energy of a video decoder. Estimation errors, which are found to be lower than $6\%$, are smaller than errors obtained using the processing time and comparable using a feature based modeling approach. Future work can make use of this knowledge to develop energy efficient decoding methods or construct energy efficient bit streams.

\bibliographystyle{IEEEbib}
\bibliography{D:/Literatur/literatureNeu}

\end{document}